\documentstyle[12pt]{article}
\begin{document}
\newcommand{\beq}{\begin{equation}}
\newcommand{\eeq}{\end{equation}}
\newcommand{\beqa}{\begin{eqnarray}}
\newcommand{\eeqa}{\end{eqnarray}}
\newcommand{\barr}{\begin{array}}
\newcommand{\earr}{\end{array}}
\newcommand{\nonum}{\nonumber}
\thispagestyle{empty}
\begin{center}
\LARGE \tt \bf{Non-Riemannian vortex geometry of rotational viscous fluids and breaking of the acoustic Lorentz invariance}
\end{center}
\vspace{2.5cm}
\begin{center} {\large L.C. Garcia de Andrade\footnote{Departamento de F\'{\i}sica Teorica - Universidade do Estado do Rio de Janeiro-UERJ.

Rua S\~{a}o Fco. Xavier 524, Rio de Janeiro, RJ

Maracan\~{a}, CEP:20550-003 , Brasil.

E-mail.: garcia@dft.if.uerj.br}}
\end{center}
\vspace{1.0cm}
\begin{abstract}
Acoustic torsion recently introduced in the literature (Garcia de Andrade,PRD(2004),7,64004) is extended to rotational incompressible viscous fluids represented by the generalised Navier-Stokes equation. The fluid background is compared with the Riemann-Cartan massless scalar wave equation, allowing for the generalization of Unruh acoustic metric in the form of acoustic torsion, expressed in terms of viscosity, velocity and vorticity of the fluid. In this work the background vorticity is nonvanishing but the perturbation of the flow is also rotational which avoids the problem of contamination of the irrotational perturbation by the background vorticity. The acoustic Lorentz invariance is shown to be broken due to the presence of acoustic torsion in strong analogy with the Riemann-Cartan gravitational case presented recently by Kostelecky (PRD 69,2004,105009). An example of analog gravity describing acoustic metric is given based on the teleparallel loop where the acoustic torsion is given by the Lense-Thirring rotation and the acoustic line element corresponds to the Lense-Thirring metric.
\end{abstract}
\vspace{0.5cm}
\vspace{2.0cm}
\newpage
\pagestyle{myheadings}
\markright{\underline{Acoustic Lorentz invariance}}

\section{Introduction}

The acoustic metric in fluids has been proposed by Unruh \cite{1} in 1981 with the purpose of investigating more realistic Hawking effect and sonic spectrum of temperature, which allowed him to proposed the concept of sonic black hole or dumb hole. More recently Garcia de Andrade \cite{2} has extended this concept to allow for acoustic torsion, with the very strong constraint of irrotational perturbations in fluids originally already in rotation to compare the Euler fluid equations with the Riemann-Cartan (RC) wave equation to obtain the so-called analog gravity models \cite{3} endowed with torsion. Actually the idea of applying non-Riemannian geometry in hydrodynamics is not new, but stands from Kazuo Kondo in 1947 \cite{4} when he used the idea of non-holonomicity to investigate hydrodynamical turbo-machines, what is new here is that torsion here is used as analogous model for rotation, distinct from Kondo case where torsion appears, for example in the Boltzmann-Hammel equation \cite{4}. Here the dynamical equations are given by the Navier-Stokes \cite{5} and conservation of mass density of the fluid equations which do not contain Cartan torsion whatsoever. In other words, the dynamics of the effective gravity here is not given by the Einstein equations, let alone by the Einstein-Cartan equations \cite{6}. Torsion only appears when we compare the fluid equations with the wave equation in the RC real spacetime. The effective or sonic torsion appears then, in terms of the parameters of the fluid such as rotation and fluid velocity. In this paper we consider the generalization of the previous paper \cite{2} to the case where viscosity appears together with rotation. One of the main advantages to include viscosity is not only to make the fluid more real (apart from superfluids which has the advantage of not possessing viscosity at all) is, as shown by M. Visser \cite{7} to be able to deal with interesting physical phenomena as acoustic Lorentz violation.
One of main differences between this previous work and the present one is that here we consider that viscosity is present partly in the acoustic torsion and in the Visser approach viscosity is present only in the acoustic metric. To take this advantage in the  present work we also show that the acoustic torsion is able to induce acoustic Lorentz violation in strong gravitational analogy with recent work by Kostelecky \cite{8} where he shows that explicitly Lorentz violation is found to be imcompatible with generic RC geometries , but spontaneous Lorentz breaking avoids this problem. By considering viscosity and fluid vorticity we also pave the way to built a non-Riemannian approach to turbulent flow. The paper is organized as follows: In section 2 we briefly review the RC wave equation of massless fields while in section 3 we present the new material concerning the non-Riemannian geometry of viscous rotational fluids and its vorticity perturbation, followed by the comparison with the RC wave equation and consequent derivation of the acoustic torsion. Section 4 presents the breaking of acoustic  Lorentz symmetry through acoustic torsion and viscosity and vorticity of the classical fluid. Section 5 contains also new material and we provide an example of teleparallel gravity torsion loop which metric can be mapped with the Lense-Thirring (LT) metric as long as the LT rotation can be associated to torsion. Since we recently show that the Letelier teleparallel loops can be associated to the metric of superfluid ${}^{4}He$ we may consider that this example is similar to the Volovik \cite{8} example of associating the LT metric to be an analog model for the superfluid model with rotation. The basic difference with Volovik's example is that his case does not contain torsion and that his metric represents the motion of phonons around the vortex in ${}^{4}He$ superfluid. In section 6 conclusions are presented. In this paper we also respond to some criticism concerning our previous work \cite{2} where irrotational perturbations would have been contaminated by the background vorticity and this would have spoiled our acoustic torsion model. This problem here does not appear since the  background flow is already rotational.  
\section{Wave Equation in Riemann-Cartan Spacetime}
In general relativistic analogue models the fluid equations are expressed in terms of the Riemannian wave equation for a scalar field ${\Psi}$ in the form
\begin{equation}
{\Box}^{Riem}{\Psi}= 0
\end{equation}
where ${\Box}^{Riem}$ represents the Riemannian D'Lambertian operator given by
\begin{equation}
{\Delta}^{Riem}=\frac{1}{\sqrt{-g}}{\partial}_{i}(\sqrt{-g}g^{ij}{\partial}_{j})
\end{equation}
In this case $(i,j=0,1,2,3)$ where g represents the determinant of the effective Lorentzian metric which components are $g_{00}=-\frac{\rho}{c}(v^{2}-c^{2})$,$g_{0i}=-\frac{\rho}{c}(v_{i})$, $g_{11}=g_{22}=g_{33}= 1$, others zero. Throughout this paper c represents the speed of sound, which quanta are given by the phonons, ${\rho}$ the fluid density, and $\vec{v}$ is the velocity of the fluid. Applying minimal coupling of torsion to the metric one obtains the covariant derivative of an arbitrary vector field $B_{k}$ in RC spacetime 
\begin{equation}
{\nabla}_{i}B_{j}={\partial}_{j}B_{j}- {{\Gamma}_{ij}}^{k}B_{k}
\end{equation}
where ${\Gamma}$ is the RC spacetime connection given in terms of the Riemannian-Christoffel connection ${\Gamma}'$ by 
\begin{equation}
{{\Gamma}_{ij}}^{k}={{{\Gamma}'}_{ij}}^{k}-{K_{ij}}^{k}
\end{equation}
where ${K_{ij}}^{k}$ are the components of the contortion tensor. These formulas allow us to write the non-Riemannian D'Lambertian as
\begin{equation}
{\nabla}_{i}{\Psi}^{i}={{\nabla}^{Riem}}_{i}{\Psi}^{i}+g^{ij}{K_{ij}}^{k}{\Psi}_{k}
\end{equation} 
To simplify future computations we consider the trace of contortion as given by $g^{ij}{K_{ij}}^{k}:=K^{k}$. We also define ${\Psi}^{i}={\partial}_{i}{\Psi}$. This definition allows us to express the non-Riemannian D'Lambertian of a scalar function as 
\begin{equation}
{\Box}{\Psi}={{\Box}^{Riem}}{\Psi}+ K^{k}{\partial}_{k}{\Psi}
\end{equation}
Thus the non-Riemannian wave equation  
\begin{equation}
{\Box}{\Psi}=0
\end{equation}
reduces to the following equation
\begin{equation}
{{\Box}^{Riem}}{\Psi}= - K^{k}{\partial}_{k}{\Psi}
\end{equation}
In the next section we shall present the dynamics of analog models through the Navier-Stokes and conservation of mass equation.
\section{Dynamics of analog models in rotational viscous fluids}
Consider the dynamics of the non-relativistic fluids by expressing the conservation of mass equation and the Navier-Stokes equation by 
\begin{equation}
{\partial}_{t}{\rho}+ {\nabla}.({\rho}\vec{v})=0
\end{equation}
\begin{equation}
(\frac{\partial\vec{v}}{{\partial}t})= \vec{v}{\times}{\vec{\Omega}}-\frac{{\nabla}p}{\rho} -{\nabla}({\phi}+\frac{1}{2}{\vec{v}}^{2})+{\nu}{\nabla}^{2}\vec{v}
\end{equation}
where ${\nu}$ is the viscosity coefficient,${\phi}$ the potential energy of the fluid and p is the pressure. Here ${\vec{\Omega}}= {\nabla}{\times}{\vec{v}}$ and $\vec{v}={\chi}{\nabla}{\Psi}$ represents the vorticity of the fluid and the vector velocity in terms of the stream function \cite{9} ${\chi}={\chi}(\vec{r)}$. Performing the perturbation to these equations according to the perturbed flow quantities 
\begin{equation}
{\Psi}= {\Psi}_{0}+ {\epsilon}{\Psi}_{1}
\end{equation}
\begin{equation}
{\vec{v}}= {\vec{v}}_{0}+ {\epsilon}{\vec{v}}_{1}
\end{equation}
\begin{equation}
{\rho}= {\rho}_{0}+ {\epsilon}{\rho}_{1}
\end{equation}
\begin{equation}
\vec{\Omega}= \vec{{\Omega}_{0}}+ {\epsilon}\vec{{\Omega}_{1}}
\end{equation}
Here we consider that both the velocity of the original fluid $\vec{v}_{0}$ and the perturbed velocity are rotational fluids. This is according to the rule that a fluid that is initially irrotationally shall remain irrotationally. The rotation
\begin{equation}
\vec{{\Omega}_{1}}={\nabla}{\times}{\vec{v_{1}}}={\nabla}{\times}{\chi}{\nabla}{\Psi}_{1}
\end{equation}
Substitution of these expressions into the conservation  equation yields
\begin{equation}
{\partial}_{t}{{\rho}_{1}}+{\nabla}.({\rho}_{1}\vec{v_{0}}+{\rho}_{0}\vec{v_{1}})=0
\end{equation}
using ${\xi}={\xi}_{0}+{\epsilon}\frac{p_{1}}{{\rho}_{0}}$ where ${\xi}=\frac{1}{\rho}{\nabla}p$, the Navier-Stokes equation becomes
\begin{equation}
-{\partial}_{t}({\nabla}{\Psi}_{1})+\frac{p_{1}}{{\rho}_{0}}-\vec{{v}_{0}}.{\nabla}{\Psi}_{1}+{\nu}{\nabla}^{2}\vec{v}-{\alpha}=0
\end{equation}
where 
\begin{equation}
{\nabla}{\alpha}:= -\vec{{\Omega}}_{0}{\times}{\chi}{{\nabla}{\Psi}} \end{equation}
which to first order in ${\epsilon}$ yields
\begin{equation}
{\nabla}{\alpha}_{1}= -(\vec{{\Omega}}_{0}{\times}\vec{v}_{1}+ \vec{{\Omega}}_{1}{\times}\vec{v}_{0})
\end{equation}
A simple vector algebra yields
\begin{equation}
{\alpha}_{1}= \int{{\nabla}{\alpha}_{1}.d\vec{r}}= -\int{(\vec{{\Omega}}_{0}{\times}\vec{v}_{1}+ \vec{{\Omega}}_{1}{\times}\vec{v}_{0}).d{\vec{r}}}
\end{equation}
Rearranging terms in the Navier-Stokes equation
\begin{equation}
p_{1}={\rho}_{0}{\chi}[{\partial}_{t}{\Psi}_{1}+\vec{v}_{0}.{\nabla}{\Psi}_{1}+\frac{{\nabla}{\alpha}_{1}}{\chi}-{\nu}{\nabla}^{2}{\Psi}_{1}]
\end{equation}
Substitution of $p_{1}$ into the conservation equation yields
\begin{equation}
-{\partial}_{t}({\chi}\frac{{\partial}{\rho}}{{\partial}p}{\rho}_{0}[{\partial}_{t}{\Psi}_{1}+\vec{v}_{0}.{\nabla}{\Psi}_{1}+\frac{{\alpha}_{1}}{\chi}-{\nu}{\nabla}^{2}{\Psi}_{1}])+{\nabla}.[{\chi}{\rho}_{0}({\nabla}{\Psi}_{1}-\frac{{\partial}{\rho}}{{\partial}p}\vec{v_{0}}({\partial}_{t}{\Psi}_{1}+\vec{v}_{0}.{\nabla}{\Psi}_{1}+\frac{{\alpha}_{1}}{\chi}-{\nu}{\nabla}^{2}{\Psi}_{1}))]=0
\end{equation}
By comparison of this equation with the RC wave equation we see that the Riemannian wave part reproduces Unruh metric while the new parts yields the following components of Cartan contortion
\begin{equation}
\vec{K}= \frac{{\rho}_{0}}{c^{2}}[{\nu}{\nabla}({\nabla}{\chi}.\vec{v}_{0})+{\chi}\vec{\Omega}_{0}{\times}\vec{v}_{0}]
\end{equation}
Note that in this formula, even if the background vorticity ${\Omega}_{0}$ vanishes the spatial part of the acoustic contortion $\vec{K}$ does not vanish, but it is proportional to the viscosity. To obtain this formula we use the vector analysis identity $\vec{v_{0}}.\vec{{\Omega}_{0}}{\times}{{\nabla}{\Psi}_{0}}=0$ since ${\vec{v_{0}}}$ is proportional to ${\nabla}{\Psi}_{0}$. The geometrical interpretation of acoustic torsion being related to viscosity and vorticity seems to be in agreement with the Cartan idea that the torsion would be not only proportional to rotation but also to the translation which would cause viscosity. The Magnus effect could also be represented by torsion since it represents a lift plus a rotation of a cylinder inside a fluid. At this point is interesting to pointed out that the apparent distinction between Visser's approach and ours could be solved if we use teleparallel geometry for acoustic torsion since in this case there will be an  equivalence between general relativity and the non-Riemannian geometrical gravity. The time component of the contortion reads
\begin{equation}
{K^{t}}{\partial}_{t}{\Psi}_{1}=\frac{{\rho}_{0}}{c^{2}}{\partial}_{t}[{\nu}{\nabla}^{2}+\vec{v}_{0}.(\vec{{\Omega}_{0}}{\times}{\nabla}+{\nabla})]{\Psi}_{1}
\end{equation}
where we have used that $\frac{{\partial}{\rho}_{0}}{{\partial}p}=\frac{1}{c^{2}}$. Note also that the contortion component $K^{t}$ is a dissipation coefficient to this equation. Throughtout the computations we have also considered the imcompressibility of the classical fluid (in this section we are considering normal part of the superfluid since the viscosity is non-zero). 
\section{Acoustic Lorentz symmetry breaking by acoustic torsion}

M. Visser \cite{8} has been able to incorporate viscosity of the flow into the acoustic metric by coupling the term ${\nabla}^{2}{\Psi}_{1}$ to the flat part of the metric. In this section we shall discuss the possibility of considering the acoustic torsion inducing acoustic Lorentz violation, taking the advantage of the presence of acoustic torsion in our formulation of the non-Riemannian geometry of viscous flow endowed with vorticity \cite{2}. Let us consider the viscous wave equation endowed with vorticity
\begin{equation}
{{\partial}_{t}}^{2}{\Psi}_{1}= c^{2}{\nabla}^{2}{\Psi}_{1}+{\partial}_{t}[{\nu}{\nabla}^{2}+\vec{v}_{0}.({\Omega}_{0}{\times}{\nabla}+{\nabla})]{\Psi}_{1}
\end{equation}
we shall now consider the eikonal approximation in the form
\begin{equation}
{\Psi}_{1}= a(x) exp(-i[{\omega}t-\vec{k}.\vec{x}])
\end{equation}
with $a(x)$ a slow varying function of position which is equivalent to ${\nabla}{a(x)}$ being approximately zero throughout the computations. We also do not consider derivatives of the metric. Thus the viscous wave equation with vorticity in the eikonal approximation reads 
\begin{equation}
{\omega}={\beta}+\frac{1}{2}{\gamma}\pm{\sqrt{{c^{2}k^{2}}-\frac{({\nu}k^{2})^{2}}{2}+{\beta}^{2}+{\beta}{\gamma}+\frac{{\gamma}^{2}}{4}}}-i\frac{{\nu}k^{2}}{2}
\end{equation}
where to simplify matters we have defined the following quantities: ${\beta}:=\vec{v}.\vec{k}$ and ${\gamma}:=\vec{{v}_{0}}.\vec{{\Omega}_{0}}{\times}\vec{k}$. From these quantities and the dispersion formula one notes that the first term is due to the bulk motion of the fluid, the second term is due to the rotational and viscosity contributions to Lorentz violation while the 
the last term is the dissipative term. Note that the Lorentz violation is supressed at low momentum. Note also that vorticity adds a term to the bulk matter contribution to ${\omega}$. Besides we note that the dissipative term does not have the contribution of vorticity.
\section{Teleparallel loops and the Lense-Thirring metric in superfluid ${}^{4}He$}
Recently we have shown \cite{2} that the Letelier metric representing a teleparallel torsion loop \cite{11} is given by the metric 
\begin{equation}
ds^{2}= (dt+\vec{B}.d\vec{r})^{2}-(dx^{2}+ dy^{2}+dz^{2})
\end{equation}
where $\vec{B}=(B_{x},B_{y},B_{z})$ is an arbitrary vector and $\vec{r}=(x,y,z)$, could be mapped to the acoustic metric of Unruh describing superfluid ${}^{4}He$. This metric can be written in terms of the basis 1-form ${\omega}^{i}$ where $(i,j=0,1,2,3)$ as
\begin{equation}
ds^{2}=({\omega}^{0})^{2}-({\omega}^{1})^{2}-({\omega}^{2})^{2}-({\omega}^{3})^{2}
\end{equation}
where ${\omega}^{0}=(dt+\vec{B}.d\vec{r})$, ${\omega}^{a}= dx^{a}$, and $(a,b=1,2,3)$. Let us now consider that torsion form is given by
\begin{equation}
T^{0}=  {\epsilon}_{abc}J_{a}{\omega}^{b}{\wedge}{\omega}^{c}
\end{equation}
Here ${\epsilon}_{abc}$ is the $3-D$ Levi-Civita totally skew symbol. Other torsion components $T^{a}=0$. From the Cartan's second structure equation 
\begin{equation}
{T^{i}}= d{\omega}^{i}+ {\omega}^{i}_{j}{\wedge}{\omega}^{j}
\end{equation}
one obtains the following \cite{2} nonvanishing connection one-form  
\begin{equation}
2{{\omega}^{i}}_{k}=-{{{{\epsilon}^{i}}_{k}}^{m}}_{n}{[{rot\vec{B}}-\vec{J}]}_{m}{\omega}^{n}
\end{equation}
Note that as pointed out by Letelier the relation
\begin{equation}
\vec{J}={\nabla}{\times}\vec{B}
\end{equation}
The connection one-forms ${{\omega}^{i}}_{k}$ would vanish on a $T_{4}$ teleparallel spacetime where the full RC curvature would vanish according to the the first Cartan's structure equation 
\begin{equation}
{R^{i}}_{j}=\frac{1}{2}{R^{i}}_{jkl}{\omega}^{k}{\wedge}{\omega}^{l}=d{{\omega}^{i}}_{j}+ {{\omega}^{i}}_{p}{\wedge}{{\omega}^{p}}_{j}
\end{equation}
Letelier \cite{11} considered that teleparallel torsion loops could be produced by choosing a torsion form as a distribution with support along a curve C with parametric equation ${\vec{x'}}={\vec{x'}({\lambda})}$ and taking the torsion vector ${\vec{J}}$ as 
\begin{equation}
{\vec{J}_{k}(x)}=I {{\delta}_{k}}^{2}(\vec{x},C)
\end{equation}
This is a two-dimensional distribution with support on the line C. The integrability condition of the curl free for $\vec{B}$ expression is ${\nabla}.\vec{J}=0$. Application of above integrability condition to Dirac distribution for torsion $\vec{J}$ expression one obtains
\begin{equation}
{\nabla.\vec{J}}(\vec{x})={\delta}^{3}(\vec{x}-\vec{x'}_{i})-{\delta}^{3}(\vec{x}-\vec{x'}_{f})
\end{equation}
Thus in order that the torsion vector $\vec{J}$ be divergent-free one must impose that the final and initial points coincides or $\vec{x'}_{i}=\vec{x'}_{f}$. In this section we show that the Letelier metric can also be mapped to the LT metric
\begin{equation}
ds^{2}=(1-\frac{{\Omega}^{2}{\rho}^{2}}{{c}^{2}})dt^{2}+2(\frac{\Omega}{c^{2}}){\rho}^{2}d{\phi}dt+\frac{1}{c^{2}}d{\vec{r}}^{2}
\end{equation}
where the LT angular velocity ${\omega}_{LT}=-{\Omega}$. It is easy to note that the Letelier and LT metric can be mapped into each other as long as we consider that 
\begin{equation}
\vec{J}={\nabla}^{*}{\times}\vec{B^{*}}
\end{equation}
where ${\nabla}^{*}=({\partial}_{t},{\partial}_{\alpha})$ and $\vec{{B}^{*}}:=(B_{t},B_{\alpha})$, where ${\alpha}=1,2$. By considering that $J^{0}:={\Omega}$ the Letelier reduced equation reduces to
\begin{equation}
\frac{{\partial}B_{\phi}}{{\partial}{\rho}}=J^{0}= {\Omega}
\end{equation}
By solving this equation we have the trick to map Letelier metric to LT spacetime given by $B_{\phi}={\Omega}{\rho}$ since in the LT metric ${\Omega}$ is constant. Thus the LT metric can describe a superfluid where the acoustic torsion is given by the LT angular velocity as demonstrated in the previous section on acoustic torsion in classical incompressible fluids with viscosity.
\section{Conclusions}
We have shown that the LT metric can be considered an acoustic superfluid metric where sonic torsion is given by the LT angular velocity. This is actually a simple example of the use of acoustic torsion in classical and quantum fluids. Future prospects include the investigation of a sonic black hole endowed with acoustic torsion and possible contributions to the acoustic Hawking effect. The acoustic Lorentz breaking due to acoustic torsion analog of vorticity and viscosity is explicitly demonstrated by making use of an eikonal geometric acoustic approximation. 
\section*{Acknowledgements}
\paragraph*{}
I would like to dedicate this paper to the memories of Professors Kazuo Kondo and Jeeva Anandan who thaught me so much about torsion and inspired me through their works and talks. I am also very much indebt to Professor W. Unruh, Dr. C. Furtado, Dr. S. Bergliaffa  and Professsor P. S. Letelier for discussions on the subject of this paper. Special thanks go to Professor V. Pagneux for some discussions on acoustic perturbations. I also thank Larry Smalley teacher and friend who told me for the first time about the applications of torsion to superfluids. Grants from CNPq (Ministry of Science of Brazilian Government) and Universidade do Estado do Rio de Janeiro (UERJ) are gratefully acknowledged.

\end{document}